\documentclass[12pt]{article}
\usepackage[latin1]{inputenc}
\usepackage[british]{babel}
\usepackage{cmap}
\usepackage{lmodern}

\usepackage{amssymb, amsmath, amsthm}
\usepackage[a4paper,top=25mm,bottom=25mm,left=25mm,right=25mm]{geometry}
\usepackage{ragged2e}

\usepackage{authblk} 
\usepackage{pifont}
\usepackage{graphicx}
\usepackage[usenames,dvipsnames,svgnames,table]{xcolor}
\usepackage[figuresright]{rotating}
\usepackage{xtab} 
\usepackage{longtable} 
\usepackage{multirow}
\usepackage{footnote}
\usepackage[stable]{footmisc}
\usepackage{chngpage} 
\usepackage{pdflscape} 
\usepackage{tocbibind} 

\usepackage{pgfplots}
\pgfplotsset{compat=1.14}
\pgfplotsset{every tick label/.append style={font=\footnotesize}}
\usepackage{setspace}

\makesavenoteenv{tabular}
\usepackage{tabularx}
\usepackage{booktabs}
\usepackage{threeparttable} 
\usepackage[referable]{threeparttablex} 
\newcolumntype{R}{>{\raggedleft\arraybackslash}X}
\newcolumntype{L}{>{\raggedright\arraybackslash}X}
\newcolumntype{C}{>{\centering\arraybackslash}X}
\newcolumntype{A}{>{\columncolor{gray!25}}C}
\newcolumntype{a}{>{\columncolor{gray!25}}c}

\newlength{\tablen}

\usepackage{dcolumn} 
\newcolumntype{.}{D{.}{.}{-1}}

\usepackage{tikz}
\usetikzlibrary{arrows, calc, matrix, patterns, positioning, trees}
\usepackage[semicolon]{natbib}
\usepackage[hyphens]{url}
\usepackage{hyperref} 
\hypersetup{
  colorlinks   = true,    		
  urlcolor     = blue,    		
  linkcolor    = blue,			
  citecolor    = ForestGreen,		  	
}
\usepackage{microtype}
\usepackage[justification=centering]{caption} 

\usepackage[labelformat=simple]{subcaption}

\DeclareCaptionLabelFormat{parenthesis}{(#2)}
\makeatletter
\renewcommand\p@subfigure{\arabic{figure}.}
\makeatother

\DeclareCaptionLabelFormat{parenthesis}{(#2)}
\makeatletter
\renewcommand\p@subtable{\arabic{table}.}
\makeatother

\def\addlegendimage{\csname pgfplots@addlegendimage\endcsname}

\usepackage{enumitem}

\setlist[itemize]{leftmargin=2.5\parindent}
\setlist[enumerate]{leftmargin=2.5\parindent}

\theoremstyle{plain}

\theoremstyle{definition}

\theoremstyle{remark}

\def\keywords{\vspace{.5em} 
{\noindent \textit{Keywords}: }}

\def\AMS{\vspace{.5em} 
{\noindent \textbf{\emph{MSC} class}: }}

\def\JEL{\vspace{.5em} 
{\noindent \textbf{\emph{JEL} classification number}: }}

\title{A note on the UEFA Champions League \\ Round of 16 draw}
\author{\href{https://sites.google.com/view/laszlocsato}{L\'aszl\'o Csat\'o}\thanks{~E-mail: \emph{laszlo.csato@sztaki.hu}} }
\affil{Institute for Computer Science and Control (SZTAKI) \\
E\"otv\"os Lor\'and Research Network (ELKH) \\
Laboratory on Engineering and Management Intelligence \\
Research Group of Operations Research and Decision Systems}
\affil{Corvinus University of Budapest (BCE) \\
Institute of Operations and Decision Sciences \\
Department of Operations Research and Actuarial Sciences}
\affil{Budapest, Hungary}
\date{\today}

\begin{document}

\maketitle

\begin{abstract}
\noindent
A paper published in a leading journal of management science has recently analysed the mechanism used for the UEFA Champions League Round of 16 draw. Since the authors have not been familiar with the related sports analytics literature, this note attempts to summarise what has already been done, and what can still be done on the issue of group draw in sports tournaments.


\AMS{90-10, 91B14}

\JEL{C44, C63, Z20}
\end{abstract}

\section{Introduction}

An article published in \emph{Management Science} \citep{BoczonWilson2022} on 2 September 2022 studies the example UEFA Champions League Round of 16 draw to call the attention of economists and market designers that transparency is a first-order concern in complex assignment problems used in practice. According to the main result, the mechanism adopted by the Union of European Football Associations (UEFA) is close to a constrained-best solution, and the organiser does not lose too much in terms of fairness by preferring transparency. 

However, \citet{BoczonWilson2022} fail to discuss the related sports analytics literature, which has resulted in some weaknesses:
\begin{itemize}
\item
The UEFA Champions League Round of 16 draw is presented as a unique field solution to a constrained assignment problem in sports. However, the same question emerges in the group draw of several competitions. Crucially, the F\'ed\'eration Internationale de Football Association (FIFA) has adopted the credible and transparent solution of the UEFA for the 2018 FIFA World Cup draw \citep{Guyon2018d} as suggested by \citet{Guyon2014a}.
\item
Previous results on the UEFA Champions League Round of 16 draw \citep{Kiesl2013, KlossnerBecker2013} and the FIFA World Cup draw \citep{Guyon2015a} have been ignored.
\item
The paper does not mention and evaluate the entirely fair (evenly distributed) algorithms proposed for the constrained assignment problem \citep{Guyon2015a, KlossnerBecker2013}.
\end{itemize}

Inspired by these issues, our note endeavours to summarise all research connected to the UEFA Champions League Round of 16 draw. Hopefully, we can also inspire future research on this topic.

\section{Is the UEFA Champions League Round of 16 draw a unique field setting?} \label{Sec2}

\citet{BoczonWilson2022} present the procedure used for the UEFA Champions League Round of 16 draw as a unique field solution to an assignment problem in sports. Here, the eight group runners-up should be matched to the eight group winners such that both the association and the group constraints are satisfied: group winners cannot play against the runner-up from the same group, or a runner-up from the same country. This setting is equivalent to a group draw with two pots, the first containing the runners-up and the second containing the group winners, where the allocation is required to meet the association and group constraints.

Then the randomisation procedure used for the UEFA Champions League Round of 16 draw is equivalent to the following mechanism:
\begin{itemize}
\item
Placing the runners-up in the eight groups according to the order in which they are drawn from their urn;
\item
Labelling the groups in alphabetical order;
\item
Placing the group winners in the first available group in alphabetical order such that any dead end---a situation when the remaining group winners cannot be assigned with satisfying all draw constraints---is avoided.
\end{itemize}
This interpretation reveals that the bipartite constraint (each pairing must be between a group winner and the runner-up) is essentially not a constraint because the draw mechanism would be entirely fair, that is, evenly distributed in the absence of the other restrictions.
Furthermore, the same randomisation procedure is used in the FIBA Basketball World Cup \citep{FIBA2019}, the FIFA World Cup draw \citep{FIFA2017c, FIFA2022a}, the draw of the European Qualifiers for the FIFA World Cup \citep{UEFA2020c}, the UEFA Euro qualifying draw \citep{UEFA2018d, UEFA2022e}, and the UEFA Nations League draw \citep{UEFA2020d, UEFA2021i}.
However, the draw rules of these competitions are more complicated than the Champions League Round of 16 draw due to the higher number of groups and teams, and---exception for the FIBA Basketball World Cup---the more complex sets of constraints.

Some papers investigate the draw mechanism of the above tournaments. 
\citet{Guyon2014a} suggests two tractable procedures for the FIFA World Cup draw that produce balanced, geographically diverse groups, and are evenly distributed, meaning that all valid assignments of the draw are equally likely. The published version, \citet{Guyon2015a} contains only one of them, however, according to \citet[p.~176]{Guyon2015a}: ``\emph{To the best of our knowledge, this is the first time that a random procedure is suggested for the final draw of the FIFA World Cup that is tractable, produces balanced groups, and satisfies the geographic constraint.}''

\citet{Csato2023i} shows how the randomisation procedure is connected to a well-known problem in computer science (generating all permutations of a given sequence), quantifies the (un)fairness of the 2018 FIFA World Cup, and evaluates the distortions caused by the draw procedure in the probability of qualification for the knockout stage for each nation.
According to \citet{Csato2023j}, a careful choice of pot labels can decrease the seriousness of uneven distribution.
\citet{RobertsRosenthal2023} aim to find mechanisms for the general constrained assignment problem that follow the uniform distribution over all feasible assignments. The authors propose two procedures using balls and bowls; both algorithms can be tried interactively at \url{http://probability.ca/fdraw/}. But they require computer draws at one or several stages, which raises the suspicion of rigging.

The FIFA World Cup draw offers an especially instructive case study. The draw of the 1990 \citep{Jones1990}, 2006 \citep{RathgeberRathgeber2007}, and 2014 \citep{Guyon2015a} World Cups were seriously unfair due to a strange policy of avoiding a group match between the unseeded South-American teams and the seeded South-American teams. Therefore, the French mathematician \emph{Julien Guyon} has suggested using the procedure for the Champions League Round of 16 draw in the FIFA World Cup draw \citep{Guyon2014a}, too (this recommendation is missing from the published version of \citet{Guyon2015a}). Fortunately, FIFA has heard the message and adopted the credible and transparent solution of the UEFA for the 2018 FIFA World Cup draw \citep{Guyon2018d}. Furthermore, the same procedure has been used for the 2022 FIFA World Cup draw---albeit, FIFA has made an annoying mistake in the allocation of the teams into pots \citep{Csato2023d}.

\section{The near-optimality of the UEFA mechanism} \label{Sec3}

The UEFA Champions League Round of 16 draw has already been investigated in 2013 \citep{Kiesl2013, KlossnerBecker2013}.
\citet{Kiesl2013} uncovers the uneven distribution of the draw procedure for the 2012/13 season and provides some uniformly distributed---but uninteresting to watch---mechanisms. Nonetheless, the author argues that there is no need to improve the randomisation procedure used by the UEFA. \citet{Kiesl2013} also proves by Hall's marriage theorem why the existence of a feasible assignment is guaranteed. The relationship between the Champions League Round of 16 draw and Hall's marriage theorem is discussed in \citet[Section~3.6]{Haigh2019}, too.

The main findings of \citet{KlossnerBecker2013}, based on the Champions League seasons played between 2008/09 and 2012/13, can be summarised as follows:
\begin{itemize}
\item
Under the currently used draw procedure, it is impossible that every feasible assignment has the same probability. Even though almost all deviations between UEFA and uniform probabilities are quite small in both absolute and relative magnitudes, even these small differences in the pairwise probabilities can change the order of the most likely opponents for certain teams (Section~3);
\item
The huge amount of money at stake translates the small probability differences into quite powerful monetary effects: in almost every season, there are teams whose expected revenue declines by more than 10 thousand euros because of the imperfect randomisation procedure used by the UEFA, while other teams unduly profit by similar amounts (Section~4);
\item
Dropping the association constraint would significantly improve the distortions of the draw (Section~5.1);
\item
An alternative mechanism, which is both fascinating for fans and able to produce the right probabilities, is provided: a random matching is generated and a fixed number of swap moves carried out in an appropriate way (Section~5.3).
\end{itemize}
However, the initial assignment of this algorithm can only be obtained by a computer draw, which could not be transparent.

Even though \citet{Guyon2014a} and \citet{Guyon2015a} discuss the FIFA World Cup draw, the proposed evenly distributed procedures based on drawing the continents first and the teams second can be easily adapted for the Champions League Round of 16 draw by drawing the countries first, followed by the teams.
These mechanisms use only balls and bowls, hence, they are entirely transparent.

As an illustration, consider the draw in the quarterfinals of the 2021/22 UEFA Champions League with three English (E), three Spanish (S), one French (F), and one German (G) club, as well as the association constraint, although that was not used according to the official rules. The number of valid assignments is 42, but drawing one of them randomly is perhaps not transparent and certainly uninteresting. On the other hand, there are only three valid allocations of countries: the F-G and three E-S pairs (6 cases), or the F/G-S, E-F/G, and two E-S pairs (18 cases each). Since the probability of the second allocation type with two E-S pairs is three times the probability of the first type with three E-S pairs (in the second type, the English/Spanish teams are not identical), the draw should be started with an urn of four balls containing the possible allocations of countries.\footnote{~If putting more balls for a given allocation type raises the suspicion of rigging, it can be avoided with fixing the English or Spanish teams at this stage such that the three balls for the second allocation type are labelled by the names of the English (or Spanish) clubs.}
Depending on the outcome of this draw, either an unconstrained assignment problem should be solved for the three English and three Spanish teams, or an unconstrained $2 \times 2$ assignment problem follows the choice of one English club (from the set of three), one Spanish club (from the set of three) and a Franch/German club (from the set of two) randomly. The procedure is fair (evenly distributed), uses a small number of bowls and balls, and allows for a nice television show of limited length.

In view of these findings, it might be misleading to state that ``\emph{although marginally better randomizations are possible, the tournament's transparency first procedure under our objective resembles the fairest possible lottery over the constrained assignments}'' \citep[p.~2]{BoczonWilson2022}, and ``\emph{the chosen procedure comes very close to achieving the fairest possible outcome}'' \citep[p.~15]{BoczonWilson2022}.

\section{The importance of constrained assignment} \label{Sec4}

According to \citet{Csato2022a}, imposing draw restrictions can be an effective tool to reduce the probability of an incentive incompatible situation. Thus, constraints can be imposed to improve competitiveness, and a remarkable trade-off exists between the number of constraints and their monetary and distortive effects. 

\section{The effect of a reversed draw order} \label{Sec5}

The third part of \citet[Proposition~2]{BoczonWilson2022} uncovers that the UEFA Champions League Round of 16 draw is asymmetric as the runners-up are drawn first. The role of the draw order in the Champions League Round of 16 draw has already been recognised by \citet[Footnote~19]{KlossnerBecker2013}. Although starting the draw with the group winners instead of the runners-up seems to have only marginal effects in the 2017/18 \citep{Guyon2017b}, 2019/20 \citep{Guyon2019d}, and 2022/23 \citep{Guyon2022c} seasons, it would be interesting to see the fairness distortions \citep[Figure~4]{BoczonWilson2022} when the group winners are drawn first because this reform has essentially no price.

\section{The implications of the association constraint} \label{Sec6}

\citet[Section~4.1]{BoczonWilson2022} thoroughly analyse the monetary and the distortive effects of the association constraint in the UEFA Champions League Round of 16 draw. Nonetheless, this draw restriction influences the outcome of the tournament, too: since the best teams are usually concentrated in some associations, they benefit from avoiding each other in the Round of 16. The implications are studied via both theoretical and simulation models in \citet{Csato2023g}.

\section{A potential field of application for the UEFA randomisation procedure} \label{Sec7}

Since the FIFA World Cup is one of the few competitions where national teams from different continents play against each other, FIFA uses a geographic constraint in order to maximise the number of these matches in the group stage. In particular, no group could have more than one team from the same qualification zone except Europe, and each group should have at least one but not more than two European teams \citep{FIFA2022a}. 

Geographic separation seems to be a reasonable criterion for world championships in all other sports. However, it is not used commonly as shown by the following examples:
\begin{itemize}
\item
In the 2019 FIBA Basketball World Cup, there was a chance of 1/4 that one group from the set of Groups A, C, E, G would contain no European teams (if Iran would have been drawn into the same group with the United States) but two of such groups would have two European teams \citep{FIBA2019}.
\item
In the 2022 IHF World Women's Handball Championship, Group G consisted of Brazil and Paraguay, although there were eight groups and only three South and Central American teams. Similarly, Congo and Tunisia played in Group F, while five groups had no African team.
\item
In the 2023 IHF World Men's Handball Championship, Group C consisted of Brazil and Uruguay, although there were eight groups and only four South and Central American teams. Similarly, Egypt and Morocco played in Group G, while four groups had no African team.
\end{itemize}
Constraints such as the geographic restriction can be easily implemented in any group draw using the randomisation mechanism of the UEFA.

\section{Conclusion} \label{Sec8}

Economists and market designers seem to be increasingly interested in analysing the theoretical properties of sports rules. They are strongly encouraged to explore what has already been done in the tournament design literature. Some survey articles \citep{KendallLenten2017, LentenKendall2021, Wright2009, Wright2014} and recent books \citep{Csato2021a, LeyDominicy2023, LeyDominicy2020} can be recommended as a starting point. Journals such as the \emph{International Journal of Sports Science \& Coaching}, the \emph{Journal of Quantitative Analysis in Sports}, the \emph{Journal of Sports Analytics}, and the \emph{Journal of Sports Economics} need also be checked to find all relevant previous research.

\section*{Acknowledgements}
\addcontentsline{toc}{section}{Acknowledgements}
\noindent
We are grateful to \emph{Julien Guyon} for useful advice.

\bibliographystyle{apalike}
\bibliography{All_references}

\end{document}